# Optical chirality in high harmonic generation

Ofer Neufeld and Oren Cohen

*Solid State Institute and Physics department, Technion - Israel Institute of Technology, Haifa 32000, Israel.*

*e-mail addresses:* ofern@tx.technion.ac.il, oren@technion.ac.il

Optical chirality (OC) - one of the fundamental quantities of electromagnetic fields - corresponds to the instantaneous chirality of light. It has been utilized for exploring chiral light-matter interactions in linear optics, but has not yet been applied to nonlinear processes. Motivated to explore the role of OC in the generation of helically polarized high-order harmonics and attosecond pulses, we first separate the OC of transversal and paraxial beams to polarization and orbital terms. We find that the polarization-associated OC of attosecond pulses corresponds to that of the pump in the quasi-monochromatic case, but not in multi-chromatic pump cases. We associate this discrepancy to the fact that the polarization OC of multi-chromatic pumps vary rapidly in time along the optical cycle. Thus, we propose new quantities, non-instantaneous polarization-associated OC, and timescale-weighted polarization-associated OC, that link the chirality of multi-chromatic pumps and their generated attosecond pulses. The presented extension to OC theory should be useful for exploring various nonlinear chiral light-matter interactions. For example, it stimulates us to propose a tri-circular pump for generation of highly elliptical attosecond pulses with a tunable ellipticity.

*Introduction.* - Optical chirality (OC), a quantity that measures the local and instantaneous density of chirality of electromagnetic (EM) waves, is a very useful concept in light-matter interactions [1]. For example, it has been used for proposing 'superchiral' fields (i.e., fields with a larger OC than circularly polarized fields) that yield ultra-dichroic interactions with chiral molecules [2,3]. However, OC has not yet been applied to chiral nonlinear optical processes. The reported results follow our motivation to explore the role of OC in the generation of helically polarized high harmonics.

In high harmonic generation (HHG) [4–6], intense ultrashort laser pulses are spectrally upconverted to the extreme UV and X-rays spectral regions. The process has been utilized for various applications, including the production of attosecond pulses [7], high resolution imaging [8–10], and probing the dynamics of electronic wave functions [11]. In HHG electrons are first tunnel-ionized, and are then accelerated by the intense laser field until they recombine with the ion and emit high harmonic radiation [5]. The trajectories of the recombining electrons can generally have durations on the same order of magnitude as that of the optical cycle of the driver, making this process non-instantaneous. In the 'standard' geometry, HHG is driven by, and results with, linearly polarized light (i.e. with zero OC). Recently, generation and applications of highly helically polarized bright high harmonics were demonstrated experimentally [12–20]. Also, HHG from chiral media was shown to be chirality sensitive [21–23]. Given these exciting developments, it is pertinent to apply the concept of OC to HHG. For example, identifying a correspondences between the OCs of the pump and high harmonic fields should lead to improved understanding and control of chiral HHG and attosecond pulses [24–29].

Here, we first divide the OC of transversal and paraxial beams to polarization and orbital terms, i.e. contributions from spin angular momentum (SAM) and orbital angular momentum (OAM). For the case where the polarization term is dominant we develop a formalism for non-instantaneous OC and apply it for analyzing helical HHG. We discover that the chirality of HHG emission driven by a multi-spectral pump corresponds to a timescale-weighted OC of the pump, which is comprised of both instantaneous and non-instantaneous chiralities. Stimulated by the new formalism, we propose a tri-circular laser field that exhibits the required dynamical symmetry for generation of circularly polarized high harmonics (just like the bi-circular field), and simultaneously it is uni-directionally chiral at all timescales (denoted a uni-chiral field). We show that such a field can produce chiral attosecond pulses, even from completely isotropic media, like helium gas. Furthermore, we show that the polarization and chirality of the attosecond pulses can be controlled by varying the timescale weighted OC of the tri-circular fields. Lastly, we identify that OC is a robust quantity for estimating the circularity of few cycle pulses, where standard definitions such as ellipticity are ambiguous.

*Polarization and orbital optical chirality.* - The OC of an EM field in vacuum is given by [1]:

$$C = \frac{\varepsilon_0}{2} \vec{E} \cdot \vec{\nabla} \times \vec{E} + \frac{1}{2\mu_0} \vec{B} \cdot \vec{\nabla} \times \vec{B} \qquad (1)$$

where we use MKS. We first derive the polarization related terms in $C$ for transversal and paraxial EM beams, denoted $C_p$. Mathematically, we neglect the z component and transverse derivatives of the EM field in Eq. (1), which yields:

$$C_p = \frac{\varepsilon_0}{2}\left(E_y\partial_z E_x - E_x\partial_z E_y\right) + \frac{1}{2\mu_0}\left(B_x\partial_z B_x - B_x\partial_z B_y\right) \quad (2)$$

Assuming the pulses have a slowly varying envelope (SVE) along the z-axis, we replace the z derivatives by time derivatives, yielding:

$$C_p = \frac{\varepsilon_0}{2c_0}\left(E_y\partial_t E_x - E_x\partial_t E_y\right) + \frac{1}{2c_0\mu_0}\left(B_y\partial_t B_x - B_x\partial_t B_y\right) \quad (3)$$

where $c_0$ is the speed of light in vacuum. Direct algebraic manipulation (see appendix A.1) leads to:

$$C_p = \frac{\varepsilon_0}{2c_0}\left|\vec{E}\right|^2\partial_t\phi(t) + \frac{1}{2c_0\mu_0}\left|\vec{B}\right|^2\partial_t\phi\left(t+\frac{\pi}{2}\right) \quad (4)$$

where $\phi = \tan^{-1}(E_y/E_x)$ is the angle of the electric field vector (the magnetic field vector is rotated by $\pi/2$). Labeling $\phi'(t)$ as the time-derivative of $\phi$, and $I = \frac{c_0\varepsilon_0}{2}\left|\vec{E}\right|^2 + \frac{c_0}{2\mu_0}\left|\vec{B}\right|^2$ the light intensity, Eq. (4) gets a compact and intuitive form:

$$C_p = \frac{1}{c_0^2}\phi'(t)\,I(t) \quad (5)$$

That is, the polarization-associated OC of paraxial beams corresponds to the product of the field's rotational velocity weighted by its intensity. This definition is an instantaneous measure for the chirality, since both $\phi'(t)$ and $I(t)$ are evaluated at time $t$. Notably, OC is a dimensional quantity that depends on the pulse envelope and frequency. Here we normalize polarization-associated OC to give a dimensionless quantity with respect to a circularly polarized pulse of a similar envelope and carrier frequency, such that for monochromatic waves the instantaneous chirality coincides with the definition of ellipticity. For example, for an EM field with a fundamental frequency $\omega$, a dimensionless envelope function $A(t)$, and a maximal amplitude $E_0$, the normalized polarization term of the OC is:

$$C_p^{\,norm} = \frac{\phi'(t)\,I(t)}{c_0\varepsilon_0\left|E_0\right|^2\omega\frac{1}{\tau}\int_0^\tau\left|A(t)\right|^2 dt} \quad (6)$$

where $\tau$ is the length of the pulse, and in the CW case the integral in the denominator vanishes to unity.

For completeness, we also derive the orbital term of the OC, denoted $C_l$. Starting with Eq. (1), assuming a beam with a SVE and neglecting the $C_p$ term leads to:

$$C_l = \frac{\varepsilon_0}{2}\left(E_x\partial_y E_z - E_z\partial_y E_x + E_z\partial_x E_y - E_y\partial_x E_z\right) + \frac{1}{2\mu_0}\left(B_x\partial_y B_z - B_z\partial_y B_x + B_z\partial_x B_y - B_y\partial_x B_z\right) \quad (7)$$

An algebraic manipulation (similar to the one leading to Eq. (4)) leads to:

$$C_l = \frac{1}{c_0}\left(I_{xz}\partial_y\phi_{xz} - I_{yz}\partial_x\phi_{yz}\right) \quad (8)$$

where we have defined planar EM field intensities: $I_{iz} = \frac{c_0\varepsilon_0}{2}(|E_i|^2 + |E_z|^2) + \frac{c_0}{2\mu_0}(|B_i|^2 + |B_z|^2)$, and planar angles: $\phi_{iz} = \tan^{-1}(E_z/E_i)$, and 'i' is the index for x/y axes. Employing the beam's paraxiality, $C_l$ is well approximated by:

$$C_l = \frac{1}{c_0}\left(I_x\partial_y\left(\frac{E_z}{E_x}\right) - I_y\partial_x\left(\frac{E_z}{E_y}\right)\right) \quad (9)$$

Equation (9) clearly represents the orbital contribution to OC, as it is non-zero for linearly polarized fields, and measures the spatial roticity of the field in the transverse plane.

We are motivated to explore the generation of helically polarized HHG and therefore consider below only the polarization term of the OC. Thus, the index "p" is dropped henceforth, and any reference to OC relates to the polarization-associated term.

*Optical chirality in HHG.* - First, we explore HHG driven by quasi-monochromatic elliptically-polarized pumps [30,31], with an ellipticity $\varepsilon$. The HHG calculations are detailed in the appendix A.2. As shown in Fig. 1, the OC of the high harmonic field corresponds well to the OC of the driving laser, even though HHG is a highly nonlinear process.

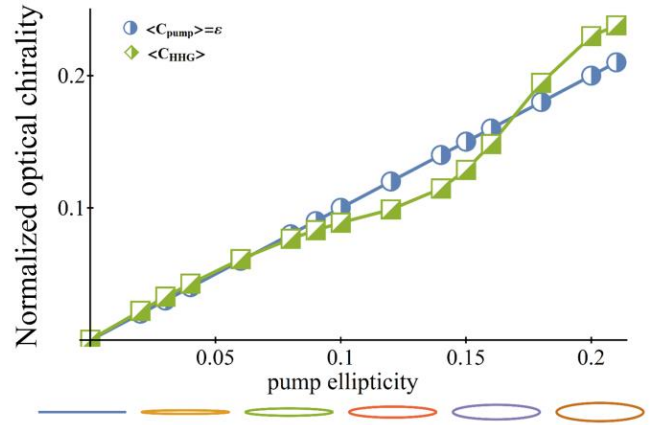

Fig. 1. Correlation of instantaneous OC of the driving monochromatic elliptical pump (which in this case is identically its ellipticity, $\varepsilon$) with numerically calculated time-averaged chirality of the emitted radiation, for $I_{max}=3\times10^{14}$ W/cm$^2$, $\omega$=800nm. Bottom shows the parametric Lissajou curve of the driver as the ellipticity is increased.

Next, we consider HHG driven by $\omega$-$2\omega$ bi-circular fields with equal amplitudes [12,32–34]. In this case, the OC of the pump varies from 0 to 0.5, and its average is $\langle C_{pump} \rangle$=0.25 (see Fig. 2(a) and (b)). The emitted attopulses on the other hand, are generally not helical [29,35,36], i.e. their OC is approximately zero. This discrepancy is not a surprise since HHG is a nonlinear non-instantaneous process, hence, it is not guaranteed that there is any correspondence between the (instantaneous) OC of the pump and HHG field. Still, motivated to correlate the OC of the HHG field with the pump chirality, we propose quantities that describe the non-instantaneous OC (the inherent duration of HHG corresponds to the time interval of the electron's motion in the continuum, which is typically in the range of ¼ to ¾ of the duration of the pump optical cycle).

*Non-instantaneous optical chirality.* - To include non-instantaneous effects for a given timescale, we alter Eq. (5) to:

$$\chi(\Delta t) = \frac{1}{c_0{}^2} \frac{\phi(t+\Delta t) - \phi(t)}{\Delta t} \bar{I}(t, \Delta t) \qquad (10)$$

where $\bar{I}(t, \Delta t)$ is an averaged intensity in times $t$ and $t+\Delta t$. Equation (10) introduces chirality by timescale, $\chi(\Delta t)$, and reduces to Eq. (5) for $\Delta t$=0. When multiple timescale are involved a weighted average is appropriate:

$$\chi_{tot} = \int_{\Delta t = 0}^{\Delta t_{\text{cut-off}}} w_{\Delta t} \chi(\Delta t) \, d(\Delta t) \qquad (11)$$

where $w_{\Delta t}$ is a weighting coefficient indicating the contribution of the timescale $\Delta t$ to the overall chirality, and $\Delta t_{\text{cut-off}}$ is a maximal cutoff for the timescales, both of which depend on the system at hand. $\chi_{tot}$ in Eq. (11) describes the timescale-weighted OC.

*Non-instantaneous optical chirality in HHG.* - We now employ the non-instantaneous chirality formalism to analyze the generation of helical attopulses driven by multi-spectral pumps.

First, we explore a counter-rotating $\omega$-$2\omega$ bi-circular pump:

$$\vec{E}_{bi}(t) = E_1 \hat{e}_R e^{i\omega t} + E_2 \hat{e}_L e^{2i\omega t} \qquad (12)$$

where $\hat{e}_{R/L}$ represents a right/left circularly polarized field vector, $\omega$ is the optical frequency related to the field's period, $T = 2\pi/\omega$, and $E_{1,2}$ are the field amplitudes. Due to its 3-fold rotational symmetry, this field generates circular high harmonics with an alternating helicity [12,16,37]. Even though the bi-circular driver produces circular harmonics, it often leads to an overall non-chiral response, and linearly-polarized attopulses [29,35,36]. Applying Eq. (10) on the bi-circular field at intensity ratios $E_1 = E_2$ (1:1), we find:

$\langle \chi(\Delta t = 0) \rangle$=0.25 and $\langle \chi(\Delta t = T/3) \rangle$=-0.5. That is, the bi-circular field at this intensity ratio changes its helicity from shorter to longer timescales, which influences the HHG process. The resulting non-chiral response can then be intuitively understood as two opposing chiral timescales that average out during the electron's motion in the continuum. This argument suggests that in order to produce a significant chiral response in the medium, the driver should be co-rotating on these timescales (thus eliminating the opposing contributions to the chirality); hence, the driver should be uni-directional at all timescales (i.e., $\chi(\Delta t)$>0 for all $\Delta t$), denoted 'uni-chiral'. One can generate such a field without breaking the discrete 3-fold dynamical symmetry by adding a third circular field at frequency $4\omega$ to the bi-circular scheme, which we denote the tri-circular scheme:

$$\vec{E}_{tri}(t) = E_1 \hat{e}_R e^{i\omega t} + E_2 \hat{e}_L e^{2i\omega t} + E_4 \hat{e}_R e^{4i\omega t} \qquad (13)$$

For $E_4 = 0$ the pump in Eq. (13) reduces to the bi-circular pump. The addition of a 4th harmonic term tilts the balance of helicity in favor of anti-clockwise rotation. Tuning the intensity ratios in Eq. (13) ($E_1 : E_2 : E_4$) allows manipulating the OC on multiple timescales. For comparison, at intensity ratios 2:1:1, the tri-circular field has a very similar shape to the bi-circular field at intensity ratios 1:1, but is uni-chiral, and rotates anti-clockwise on all timescales (Fig. 2a and 2c).

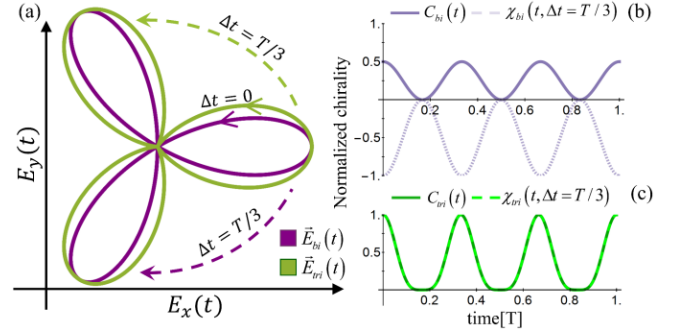

Fig. 2. Optical chirality of bi-circular and tri-circular fields, at intensity ratios 1:1 and 2:1:1, respectively. (a) Lissajou curve of the bi-circular (purple) and tri-circular (green) fields. Full arrows represent the instantaneous motion of the field, while dashed arrows represent motion on T/3 timescales. The instantaneous and long motions are uni-directional for the tri-circular field while they have opposite directions in the bi-circular fields. (b) Instantaneous and non-instantaneous (for $\Delta t = T/3$) optical chiralities of bi-circular pump. (c) Same as (b), but for tri-circular pump, where $C_{tri}(t)$ and $\chi_{tri}(t, \Delta t = T/3)$ coincide.

We numerically show that this configuration generates a chiral response from an isotropic medium on a single atom level from an initial 1s state (see appendix A.2 for numerical details). Filtering out below ionization potential (Ip) harmonics yields a highly helical attopulse train (Fig. 3(b)) with an averaged OC of 0.64 (equivalent to an ellipticity of 0.66), as opposed to the bi-circular field that generates a linearly polarized attopulse train with an averaged OC of 0.08 (Fig. 3(a)). From a spectral point of view, even though the driver generates both left and right circular harmonics, the intensity of each right rotating harmonic is about five times larger than that of the nearby left rotating harmonic. The main difference between the driving fields is the long-term rotation directionality, indicating that this non-instantaneous chirality of the pump leads to the drastic change in the emitted radiation's chirality. The drawbacks for adding the $4\omega$ field are reductions in conversion efficiency (by an order of magnitude) and cutoff energy.

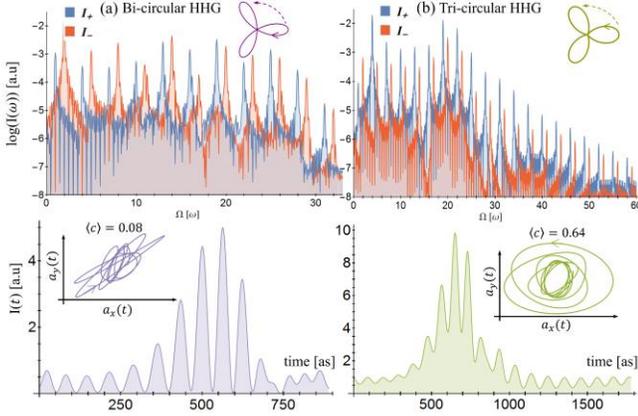

Fig. 3. Numerical HHG spectral intensity projected onto left and right rotating components in log scale (top) and emitted attopulse trains (bottom) from: (a) bi-circular $\omega$-$2\omega$ driver with intensity ratios 1:1, $I_{max}=2 \times 10^{14}$ W/cm², $\omega$=800nm, (b) uni-chiral tri-circular $\omega$-$2\omega$-$4\omega$ driver with intensity ratios 2:1:1, $I_{max}=3 \times 10^{14}$W/cm², $\omega$=1600nm. Top inset shows a Lissajou curve of the driver, and bottom inset a Lissajou curve of a single burst of the emitted attopulse train.

Next, we investigate a wider range of intensity ratios in the tri-circular scheme and examine the correspondence between the chirality of light emitted from HHG to the chirality of the driver. We define a parameter $\eta$ which is varied from 0 to 1 as the tri-circular field changes its intensity ratios from 1:1:0, to 2:1:1 The pump then has the form:

$$\vec{E}_{tri,\eta}(t) = E_{0,\eta}\left[(1+\eta)\hat{e}_R e^{i\omega t} + \hat{e}_L e^{2i\omega t} + \eta \hat{e}_R e^{4i\omega t}\right] \quad (14)$$

where $E_{0,\eta}$ is normalized per $\eta$ to keep the peak amplitude constant. The spatiotemporal shape of this field is shown for several values of $\eta$ in the bottom of Fig. 4. For each value of $\eta$ we calculate the OC of the emitted radiation. Fig. 4 clearly shows that varying the intensity ratios in the driver tunes the chirality of the emitted light. Also, from Fig. 4 one can tell that the instantaneous OC of the pump generally does not correspond to the instantaneous OC of the high harmonic waves. For example: for $\eta$=0.3 the time-averaged OC of the driver is $\langle C_{pump} \rangle$=0, while the emitted light is chiral ($\langle C_{HHG} \rangle$>0). Also, for $\eta$=0.2, the OC of the driver is negative ($\langle C_{pump} \rangle$<0), while that of the emitted light is positive

($\langle C_{HHG} \rangle$>0). Assuming that the $\Delta t$=$T/3$ timescale is important, we approximate the total chirality of the driver with Eq. (11) as comprised from just two dominant timescales, $\Delta t$=0 and $\Delta t$=$T/3$:

$$\langle \chi_{tot} \rangle = w_0 \langle \chi(0) \rangle + w_{T/3} \langle \chi(T/3) \rangle \quad (15)$$

Under this assumption we find a correspondence between the timescale weighted OC of the pump to the instantaneous OC of the HHG field. A best-fit is found for $w_0$=2.14$w_{T/3}$, with $R^2$=0.993, as seen in Fig. 4. We can also gain some physical intuition on the system at hand from the weighting ratios of the two timescales: in tri-circular HHG the instantaneous timescale is more significant than the $T/3$ timescale. Similar results are also obtained for a 4-fold symmetric tri-circular pump with frequencies $\omega$-$3\omega$-$5\omega$, where the T/4 timescale plays a significant role (see appendix A.3).

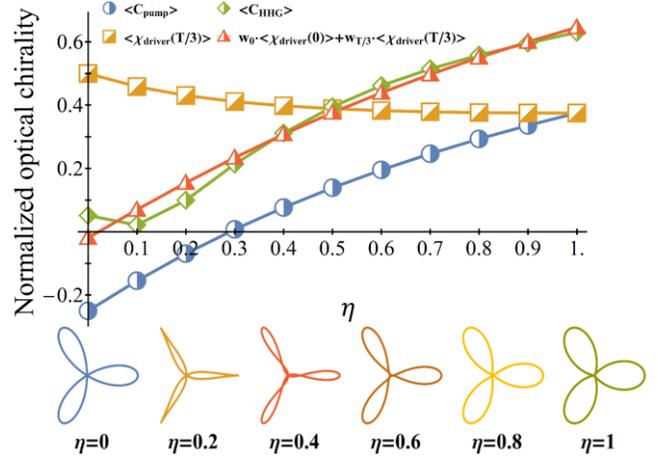

Fig. 4. Correspondence between time-averaged non-instantaneous OC of the driving tri-circular pump to the numerically calculated time-averaged OC of the emitted radiation, for $I_{max}$=3 × 10¹⁴ W/cm², $\omega$=1600nm. Bottom shows the parametric Lissajou curve of the driver as $\eta$ is varied. The best fit using two timescales is obtained for $w_0$ = 2.14$w_{T/3}$, with an $R^2$ = 0.993 between the green and red curves.

We also applied the OC approach for investigating attopulses generated by the bi-circular scheme with varying amplitude ratios. This scheme was proposed and implemented experimentally to produce highly chiral overall HHG spectra [29], which should also correspond to attosecond pulses with large ellipticity (the first experimental highly chiral overall HHG spectra using bi-circular pumps was reported in ref. [19]). In ref. [29], the intensity ratios in the counter-rotating bi-circular field were tuned such that $I_\omega$>$I_{2\omega}$, which results in highly chiral attopulses. A symmetric manipulation of $I_{2\omega}$>$I_\omega$ leads to less chiral attopulses. The non-instantaneous OC theory elucidates that this occurs because as for $I_\omega$>$I_{2\omega}$ one produces a uni-chiral pump, while for $I_{2\omega}$>$I_\omega$ the pump's instantaneous chirality is increased but remains opposite to its non-instantaneous chirality, resulting in a reduced chiral response (Fig. 5).

Interestingly, oscillations in the OC of emitted attopulses in Figs. 4 and 5 appear when the instantaneous and non-instantaneous chiralities of the pump are oppositely signed (the oscillations are larger in Fig. 5 because the difference between the instantaneous and non-instantaneous chiralities is larger).

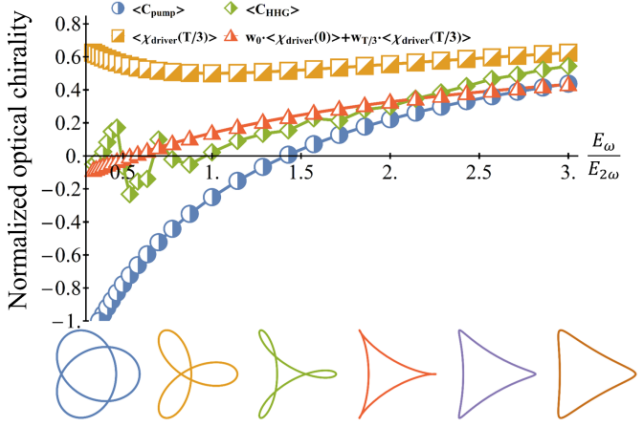

Fig. 5. Same as in Fig. 4, but for a bi-circular pump (Eq. (12)) with varying amplitude ratios between its first and second harmonic, for $I_{max}$=2.5 × $10^{14}$ W/cm², $\omega$=800nm. OC is calculated for emitted attopulses (below Ip harmonics are filtered out).

Lastly, we note that a time-averaged OC ($\langle\langle C\rangle\rangle$) provides a robust and effective quantity for estimating an EM field's circularity, where standard definitions such as ellipticity can be ambiguous. Ellipticity becomes ambiguous if pulses are comprised of broadband spectra (such as in HHG), and the spatiotemporal profile of the pulse no longer resembles an ellipse. For instance, using stokes parameters [38] to calculate the 'ellipticity' of a $\omega$-$2\omega$ counter-rotating bi-circular field (at amplitude ratios 1:1) results in an ellipticity $\varepsilon$=1, even though the field is clearly not 'as circular' as circularly polarized light. Using OC gives a more realistic estimate of $\langle C\rangle$=0.25, i.e. still circular, but less than circularly polarized light. This issue naturally also occurs in more complex wave forms that are comprised of multiple high harmonics, thus we propose using the averaged OC to evaluate the circularity in these cases.

*Summary.* - We presented a non-instantaneous optical chirality theory, and used it to analyze multi-chromatic helical HHG. We first divided the OC of transverse and paraxial beams to polarization and orbital terms, and then extended polarization-associated OC to include non-instantaneous contributions. Our results show that the chirality of the emitted HHG field corresponds to both the instantaneous and non-instantaneous chiralities of the pump. We use this intuition to predict 'uni-chiral' tri-circular field configurations that are uni-directionally helical, and can drive highly helical attopulse trains from an isotropic medium. Moreover, the attopulses' polarization state can be controlled by tuning the timescale weighted chirality of the pump. Lastly, we recommended the use of averaged OC to evaluate the circularity of broadband ultrashort pulses.

This work paves the way to various new research directions. First, application of OC in other nonlinear optical processes may yield new insights. Second, the separation of OC to polarization and orbital terms allows investigating regimes where the terms are of similar magnitude and might interact, or convert from one another. Third, the introduction of uni-chiral fields to HHG could lead to enhanced selectivity of chiral HHG spectroscopy, and an improved understanding of nonlinear chiral light-matter interactions [21,22]. Four, OC theory is also applicable to strong-field ionization of atoms and molecules, and could be useful for producing and controlling electron vortices, rotational electron currents, and spin-polarized electrons [39–42]. Lastly, non-instantaneous OC and uni-chiral fields should prove useful for the production of intense ultrashort magnetic field pulses [43].

*Acknowledgements.* - This work was supported by the Israel Science Foundation (grant No. 1225/14), the Israeli Center of Research Excellence 'Circle of Light' supported by the I-CORE Program of the Planning and Budgeting Committee and the Israel Science Foundation (grant No. 1802/12), and the Wolfson foundation. O.N. gratefully acknowledges the support of the Adams Fellowship Program of the Israel Academy of Sciences and Humanities.

*Appendix A.1: transition from Eq. (3) to Eq. (4).* - The transition from Eq. (3) to Eq. (4) uses the following identity, both on the electric and magnetic fields:

$$\left(F_y\partial_t F_x - F_x\partial_t F_y\right)$$
$$=\left(F_x{}^2 + F_y{}^2\right)\frac{1}{1+\left(\dfrac{F_y}{F_x}\right)^2}\frac{\left(F_y\partial_t F_x - F_x\partial_t F_y\right)}{F_x{}^2} \qquad (16)$$
$$\equiv\left(F_x{}^2 + F_y{}^2\right)\frac{\partial}{\partial t}\left[\tan^{-1}\left(\frac{F_y}{F_x}\right)\right]$$

where $\vec{F}$ is a time-dependent vector.

*Appendix A.2: HHG numerical details.* - Numerical calculations of the high harmonic spectrum were performed by solving the time-dependent 2D Schrödinger equation in the length gauge, within the single active electron approximation, and the dipole approximation. The system's time-dependent Hamiltonian is given in atomic units by:

$$\mathcal{H}(t) = \left( -\frac{1}{2}\vec{\nabla}^2 + V_{atom}(\vec{r}) + \vec{r} \cdot \vec{E}_{pulse}(t) \right) \quad (17)$$

where $V_{atom}$ represents a spherically symmetric coulomb softened atomic potential well, set to describe the ionization potential of Ne ($I_p = 0.793\ a.u.$) [26]:

$$V_{atom}(\vec{r}) = -\frac{1}{\sqrt{\vec{r}^2 + 0.1195}} \quad (18)$$

$\vec{E}_{pulse}(t)$ is the electric field of the pump pulse, defined by the electric field of monochromatic elliptical, or multi-chromatic pumps as specified in the main text, respectively, times a flat-top envelope function with a 4 fundamental cycle long rise and drop sections and a 6 cycle long flat-top. The initial wave function was chosen as the atomic 1s ground state found by complex time propagation. The Schrödinger equation was discretized on a square real-space grid of size $L \times L$ for $L$=120a.u., with spacing $dx=dy$=0.2348a.u., and propagated with a 3$^{rd}$ order split operator method [44,45] with a time step $dt$=0.01a.u. Convergence was tested with respect to grid size, density, and time-step. Absorbing boundaries were used with the absorber set to (in a.u.):

$$V_{ab}(\vec{r}) = -i5 \times 10^{-4} \Theta\left( \left(|\vec{r}| - 36\right)^3 \right) \quad (19)$$

where $\Theta$ represents a Heaviside step function. The dipole acceleration was calculated using Ehrenfest theorem [46], from which the harmonic spectrum is found by Fourier transform. The OC of the HHG field is calculated after removing the fundamental harmonics in the spectrum, and is normalized with respect to the OC of the most chiral attopulse train emitted from the given geometry (after filtering out below Ip harmonics).

*Appendix A.3: 4-fold tri-circular pumps.* - This appendix presents a similar analysis as that in the text for tri-circular $\omega$-$2\omega$-$4\omega$ pumps, but for a 4-fold symmetric tri-circular field comprised of the $\omega$-$3\omega$-$5\omega$ frequencies. This arrangement allows manipulating the chirality of the pump by tuning the intensity ratios between the different colors without breaking the 4-fold symmetry. We define the same parameter $\eta$ which is varied from 0 to 1, and the pump field varies as follows:

$$\vec{E}_{tri,\eta}^{4-fold} = E_{0,\eta}\left[ (1+\eta)\hat{e}_R e^{i\omega t} + \hat{e}_L e^{3i\omega t} + \eta\hat{e}_R e^{5i\omega t} \right] \quad (20)$$

The results are seen in Fig. A.3, which shows a correspondence between the HHG field's instantaneous OC and both instantaneous and non-instantaneous contributions to the chirality, dominated by the T/4 timescale.

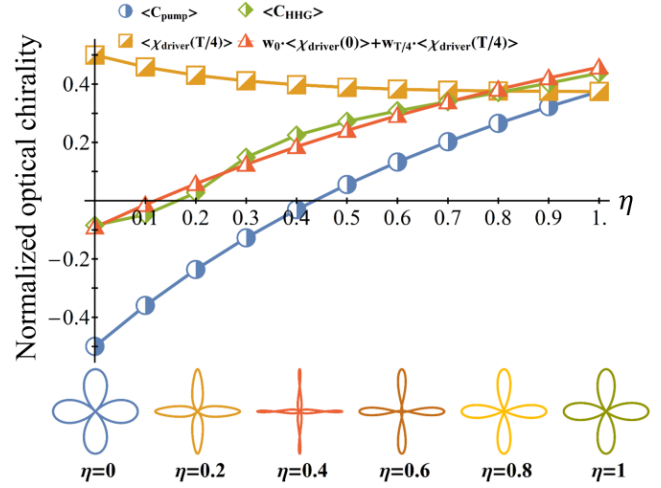

Fig. A.3. Same as in Fig. 4, but for a tri-circular 4-fold symmetric field (Eq. (20)), for $I_{max} = 3 \times 10^{14}$ W/cm$^2$, $\omega$=1600nm.